\documentstyle[sprocl,epsfig,graphics]{article}

\def\be{\begin{equation}}
\def\ee{\end{equation}}
\def\bea{\begin{eqnarray}}
\def\eea{\end{eqnarray}}
 \newcommand{\ket}[1]{|\kern.3ex#1\kern.3ex\rangle}
 \newcommand{\bra}[1]{\langle\kern.3ex #1 \kern.3ex|}
 \newcommand{\mean}[1]{\langle#1 \rangle} %moyenne statistique
 
 \newcommand{\APPROX}[1]{
   {{\raisebox{-.3cm}{$\textstyle\simeq$}} \atop {\scriptstyle{#1}}}}

 \def\re{\mbox{\rm Re}\:} %partie reelle
 \newcommand{\Tr}[1]{\mbox{\rm Tr}\left\{ #1 \right\} }  %trace

\begin{document}

\title{HALL CONDUCTIVITY FOR TWO DIMENSIONAL MAGNETIC
SYSTEMS\footnote{IPNO-TH 96-51}}
\author{ J. DESBOIS, S. OUVRY and  C. TEXIER}

\address{Division de Physique Th\'eorique\footnote{Unit\'e de Recherche des
Univresit\'es Paris 11 et Paris 6 associ\'ee au CNRS}, IPN, \\91406 Orsay, France}


%%%%%%%%%%%%%%%%%%%%%%%%%%% ABSTRACT %%%%%%%%%%%%%%%%%%%%%%%%%%%%%%%%%%%%%

\maketitle\abstracts{A Kubo inspired formalism is proposed to compute the longitudinal
and transverse
dynamical conductivities of  an  electron
in a plane (or a gas of electrons at zero temperature) coupled to the potential
vector of an external local magnetic field, with the additional coupling of the spin degree of freedom of the
electron to the local magnetic field (Pauli Hamiltonian). As an
example, the
homogeneous magnetic field Hall conductivity  is rederived. The
case of the vortex at the origin is worked out in detail.
This system  happens to
display a transverse Hall conductivity ($P$ breaking effect) which is subleading in volume
compared to the homogeneous field case, but diverging at small frequency
like $1/\omega^2$. A
perturbative  analysis is proposed for the conductivity in the
random magnetic impurity problem (Poissonian vortices in the plane). At
first order in perturbation theory, the Hall conductivity displays
oscillations close to the classical straight line conductivity of the mean
magnetic field.}

PACS: 05.30.-d; 05.60.+w; 11.10.-z
%%%%%%%%%%%%%%%%%%%%%%%%%% INTRODUCTION %%%%%%%%%%%%%%%%%%%%%%%%%%%%%

\section{Introduction : A classically free system}

Consider~\cite{nous} a Poissonian distribution of infinitely  thin vortices carrying a flux
$\phi$, perpendicular to a plane (herafter called magnetic impurities). One would like to ask about the way
 such a system could be equivalent to  a homogeneous mean magnetic field,
 or, on the contrary, how it could localize electron wavefunctions, in such
 a way that their conductivity is  altered. This question might seem
 academic at first sight. However, it is commonly believed~\cite{jansen} that in
 Quantum Hall devices  disorder does  play a crucial
 role in the understanding of plateaus for the Hall conductivity  as a
 function of $1/B$ or $N(E_F)$, the number of electrons, at
 integer (or fractional) values in unit of 
 $e^2/h$. In the case of a homogeneous $B$ field, linear response of
 the system to a small electric field gives
 no hint of such a remarkable behavior,
 since all states are delocalised and have the same
 transverse conductivity which
 varies linearly with $1/B$ or $N(E_F)$ 
(classical straight line). Disorder is needed to explain why some states
 (in fact most of them) are localised in broadened Landau levels, thus
  the plateaus in the Hall conductivity, around the classical line.
 The random magnetic impurity model 
is a model where the disorder is
 contained in the definition of the magnetic field itself~\cite{autre}, and not added as
 an external random potential. 
After averaging over disorder,  the model is translation invariant, as
the usual Landau system.
Translation invariance is at the origin of the
infinite degeneracy of Landau levels, which in turn 
is at the root of the understanding
of the macroscopic value of 
the Hall conductivity. We believe that the random magnetic impurity model
encompasses at the same time generic features of a true Landau system (and in
a certain limit both systems are indeed equivalent) and of a
disordered system. 
Finally, it is in essence a topological model, without any classical
counterpart. But 
topological considerations~\cite{thouless} have
been put forward in the past 
to explain why, in the Quantum Hall Effect, although most of the states are 
in principle localised, still the quantized
Hall conductivity remains
macroscopic.

It has been shown~\cite{nous} that two
phases coexist in the random magnetic impurity model: a
phase where the average density of states of a test particle (an
electron) coupled to the random magnetic
impurity distribution  exhibits Landau level oscillations, 
and a phase with no oscillation. We refer
to the former phase as an ordered phase, since the test particle "sees" a
mean magnetic field irrespectively of disorder, and to the latter as a
disordered phase, where the test particle "sees"
impurities individually, and  where disorder plays an important role.
  In the thermodynamic limit, the system is described by two parameters,
 $\alpha=\phi/\phi_o$, where $\phi_o$ is the flux quantum of the electron, and $\rho$,
 the mean impurity density. The natural interval of definition of $\alpha$
 is $[-1/2,1/2]$, since the system is unchanged when the fluxes are
 shifted by a multiple of the quantum of flux $\alpha\to\alpha+1$. 
 In the absence of a physical orientation to the plane, as
 coming from an exterior
 magnetic field for example, reversing the sign of the flux
 $\alpha\to-\alpha$ does not change
 the system either. So, one can restrict the interval of definition of
 $\alpha\in [0,1/2]$. Note that   the periodicity and  the symmetry
 $\alpha\to -\alpha$ in the spectrum imply that the partition function
 is unchanged when  $\alpha\to 1-\alpha$, and therefore depends only on
 $\alpha(1-\alpha)$. For a Poissonian distribution, with magnetic impurities
 randomly dropped on the plane without correlations -the simplest distribution
 considering the impurities as a Bose gas at zero temperature and 
 introducing no additional parameter in the model-, one finds that 
 the partition function of the mean magnetic field 
 $e\mean{B}=h\rho\alpha$ is reproduced as a series
 in power of $(\rho\alpha)^n$, i.e. in terms of the average problem at $n$
 impurities. This result is  a little surprising  since it amounts to say
 that a purely free classical system -an electron coupled to Aharonov-Bohm
 fluxes- leads, when quantized, to the  quantum Landau problem. Strictly speaking, this
 limit
 is attained when $\alpha\to 0, \rho\to \infty, \rho\alpha$ fixed, which is
  singular. It is possible  to organize in a systematic way 
 perturbative corrections to the leading 
 mean magnetic  field expansion.  It is
 sufficient to go at order 
 $\rho^2\alpha^4$, i.e. an electron interacting 4 times with 2 impurities, 
 to get a critical value $\alpha_o=0.28$ below which one is certain that the
 average density of states has Landau level oscillations. 

What we are presently interested in is the transport properties of an
electron, or a gas of electrons at zero temperature,
coupled to  random magnetic impurities. We know for sure that 
in the mean magnetic field limit the conductivity is the Landau
conductivity (classical straight line for the transverse conductivity).
We would like to see if the
corrections due to disorder  are going to produce oscillations -plateaus?- 
around
the classical straight line.
The paper is organized as follows: in the first section we will present a
Kubo inspired formalism~\cite{jansen}  where the dynamical linear response of a general 
2d Pauli magnetic system to a small 
external electric  field will be expressed in terms of the propagators of
the unperturbed Hamiltonian. We will in 
particular emphasize that the time derivative of the 
dynamical conductivity is easier to compute 
than the conductivity
itself, due to the particular structure of the Pauli Hamiltonian.
In the following section we will review, as a warm up
exercise,  the case of a homogeneous magnetic field (Landau problem), 
examine the difficulties inherent to the translation
invariance of the system, and the zero width of the
Landau levels, and present a way to circumvent them. Next, we
will consider two examples of magnetic systems
 where the conductivity can be exactly
computed : i) one electron or a gas of electrons coupled to one vortex, ii) 
one electron or a gas of electron coupled
to a homogeneous magnetic field and one vortex.  These systems are of
particular interest
since they both are related to the random magnetic impurity problem. In
the next section, we will consider the random
magnetic impurity problem itself, and use perturbative methods to get
information on the Hall conductivity of this system.
Finally, in the last section a discussion will follow where general conclusions will  be drawn
from the results obtained.

%%%%%%%%%%%%%%%%%%%  GENERAL FORMALISM %%%%%%%%%%%%%%%%%%%%%%%%%%%%%%%%%

\section{Conductivity  for Pauli magnetic systems}

By magnetic systems we mean in  general the class of
Hamiltonians for an electron minimally
coupled to a vector potential
$\vec A(\vec r)$ with the additional coupling of the electron spin up or down
$\sigma_z=\pm 1$ to
the local magnetic field $B(\vec r)$ (we set the electron mass $m_e=\hbar=1$)
\be
\label{AC} 
H={1\over 2}\left(\vec{p}-e \vec{A}(\vec r)\right)^2  
- {eB(\vec r)\over 2}\sigma_z
\ee
It rewrites
\be \sigma_z=+1\quad H_u={1\over 2}\Pi_-\Pi_+\ee
\be \sigma_z=-1 \quad H_d={1\over 2}\Pi_+\Pi_-\ee
where $\Pi_{\pm}=(p_x-eA_x)\pm i(p_y-eA_y)$ are the covariant momentum operators.
In the homogeneous field case, the spin coupling is  a trivial
constant shift, but, in general, it has important effects. In the 
one vortex or magnetic impurity cases, it is a sum of
$\delta(\vec r-\vec r_i)$ functions, which is needed to define in a 
non ambiguous way~\cite{lozano}~\cite{nous'}  the short
distance behaviour of the wavefunctions at the location of the impurities 
$\vec r_i$.

One would like  to evaluate the linear response of such  systems to a
small homogeneous external electric field in the $\vec x$ direction, i.e. to compute the
longitudinal and transverse local 
conductivities $\sigma_{xx}(\vec r,t)$ and $\sigma_{yx}(\vec r,t)$.
Conductivities characterize the non equilibrium dynamics of the system
under the influence of an electric field. Linear response theory   relates
them to the equilibrium
correlation functions of the Hamiltonian $H$. In a time dependent formalism, one
can start with  an  electric field  $\vec E=\delta (t)\vec
E_o$, a single impulsion at initial time,
to get a local current 
\be {\vec j}(\vec r)={e\over 2}\{\vec v \ket{\vec r}\bra{\vec r}+
\ket{\vec r}\bra{\vec r}
\vec v\}\ee
($\vec v$ is the velocity operator $\vec v=\vec p-e\vec A$) proportional to the conductivity 
\begin{equation}\label{def}
\sigma_{ij}(\vec r,t)=i\theta(t) e \Tr{\rho[j_i(\vec r,t),r_j\,]}
\end{equation}
where $\theta(t)$ is the Heaviside function.  $\Tr{\rho\cdots}$ is 
the thermal Boltzmann or Fermi-Dirac average. It can as 
well mean expectation value in a given quantum state if one wishes to 
compute the conductivity for an electron in this  state, in 
which case $\rho$            is  the projector on this quantum state.
$j_i(\vec r,t)$ is the current density operator in the Heisenberg
representation
\begin{equation}
\vec{j}(\vec r,t)=e^{i H t}{\vec j}(\vec r) 
e^{-i H t}
\end{equation}

We find convenient to compute  conductivities in a propagator formalism. 
The propagator is defined as
\begin{equation}\label{propag}
G_{\beta}(\vec{r},\vec{r}\,')= \bra{\vec{r}} e^{-\beta H}\ket{\vec{r}\,'} 
= \sum_{n} \varphi_n(\vec{r})\varphi_n^*(\vec{r}\,')\,e^{-\beta E_n}
\end{equation}
where $\{\ket{\varphi_n}\}$ is a complete set of eigenstates for the
Hamiltonian $H$.
In the case of a thermal average for  Boltzmann statistics,
$\rho$ is the usual density operator $\rho=\exp(-\beta H)/Z_{\beta}$, with
\be Z_{\beta}=\int d\vec r \:G_{\beta}(\vec r,\vec r)\ee
the partition
function.

Consider the linear combinations 
\begin{equation}\label{condcompl}
\sigma^+(\vec r,t) = \sigma_{xx}(\vec r,t) + i\sigma_{yx}(\vec r,t),\quad
\sigma^-(\vec r,t) = \sigma_{xx}(\vec r,t) - i\sigma_{yx}(\vec r,t)
\end{equation}
The local thermal (Boltzmann) conductivity, $(\sigma\equiv \sigma^{\pm})$
\be \sigma_{\beta}(\vec r, t)=\sum_n\sigma_n(\vec r,t)
e^{-\beta E_n}/Z_{\beta}\ee
where  
$\sigma_n(\vec r, t)$ is the local conductivity 
for one electron in the state $\ket{\varphi_n}$,
rewrites in terms of propagators
\begin{eqnarray}\label{local}
\sigma_{\beta}^{\pm}(\vec r,t)&=&
 {1\over 2}i\theta(t)e^2\frac{1}{ Z_{\beta}}\int d\vec{r}\,'
\Big(\Pi_{\pm} G_{i t}(\vec{r},
\vec{r}\,')x' G_{\beta-i t}(\vec{r}\,',\vec{r})\nonumber\\
&+& 
G_{i t}(\vec{r},\vec{r}\,')x'(\Pi_{\pm}^{\dagger})^* G_{\beta-i t}(\vec{r}\,',\vec{r}) - (i t\ \rightarrow\ i t+\beta)\ \Big)
\end{eqnarray}
where $\Pi_\pm^{\dagger}$ is the hermitian conjugate of $\Pi_\pm$. In
(\ref{local}) and in all formulae that follow, the differential operators $\Pi$
are always understood as acting on the $\vec r$ variable of the propagator
that immediately follows them.

The local conductivity $\sigma_\beta(\vec r,t)$ leads  upon space integration
to the volume average global conductivity, for the global current density
$\vec j=e \vec v/V$,
\be\label{pi1}
\sigma^{\pm}(t)=i\theta(t)\ \frac{e^2}{V} \Tr{ \rho[\Pi_{\pm}(t),x] }
\ee
In terms of propagators it rewrites as
\bea
\label{thermal}
\sigma_{\beta}^{\pm}(t)&\equiv& {1\over V}\int d\vec r\, \sigma^{\pm}_{\beta}(\vec r,t) \\
&=&
 i\theta(t)\frac{e^2}{V}{1\over  Z_{\beta}}\int d\vec{r}d\vec{r}\,'
\Big(\Pi_{\pm} G_{i t}(\vec{r},
\vec{r}\,')x' G_{\beta-i t}(\vec{r}\,',\vec{r}) -
        \ (i t\ \rightarrow\ i t+\beta)\ \Big) \nonumber
\eea
In the case of a system invariant by translation, the local conductivity $\sigma_\beta(\vec r,t)$ does 
not depend on $\vec r$, and both terms in  (\ref{local}) give 
the same contribution. 

To deduce from (\ref{thermal}) the conductivity of a gas of
electrons at zero temperature and Fermi energy $E_F$, one uses the
integral representation of the step function $\theta(E_F-H)$
\be \label{fermi}
\sigma_{E_F}(t)=\lim_{\eta',\epsilon'\to 0^+}
\int_{-\infty}^{\infty}\frac{dt'}{2i\pi}{e^{iE_Ft'}\over
t'-i\eta'}Z_{\beta\to it'+\epsilon'}\sigma_{\beta\to it'+\epsilon'}(t)
\ee
where $\epsilon'$ and $\eta'$ are regulators which have to be set to zero at 
the end.
The second-quantized conductivity rewrites as a sum
of first quantized conductivities  because, as 
defined in    (\ref{def}), it is a thermal average of a commutator of second quantized
operators which are sums of first quantized operators.
In the sequel we will refer to $\sigma_{E_F}$ as given in (\ref{fermi}) as the "Fermi transform" of 
$\sigma_{\beta}$. One can rewrite (\ref{fermi}) as
\be\label{dE} {d\sigma_E(t) \over dE}={1\over 2\pi }
\int_{-\infty}^{\infty}dt'{e^{iEt'}} 
Z_{\beta\to it'+\epsilon'}\sigma_{\beta\to it'+\epsilon'}(t)
\ee
with 
\be \sigma_{E_F}(t)=\int_0^{E_F}{d\sigma_E(t)\over dE} dE\ee
In (\ref{dE}), the integration starts at $E=0$ since the spectrum is
bounded from below at zero energy
(Pauli Hamiltonian (\ref{AC})). 

In practice (Quantum Hall Effect devices), the number of electrons $N({E_F}) $ 
may be fixed, implicitly determining $E_F$ by
\be 
N({E_F})= \frac{1}{2\pi i}\int_{-\infty}^{\infty}dt'\frac{e^{iE_Ft'}}{t'-i\eta'}
Z_{it'+\epsilon'}
\ee
We also have for the density of states $\rho(E)$
\be\label{rho}
\rho(E)\equiv {dN(E)\over dE}=\frac{1}{2\pi }\int_{-\infty}^{\infty}dt'e^{iEt'}
 Z_{it'+\epsilon'}
\ee
with accordingly
\be N(E_F)=\int_0^{E_F}{dN(E)\over dE}dE\ee
In general, it will be more convenient
to calculate the derivative $\dot\sigma(t)$  of $\sigma(t)$ with respect to time, rather than
$\sigma(t)$ itself. In the case
of the thermal Boltzmann conductivity, one gets
\begin{equation}\label{time}
\dot\sigma_{\beta}^{\pm}(t)=\frac{e^2}{V}\,\delta(t)  \pm  \theta(t)\,
\frac{e^2}{V}{1\over Z_{\beta}}\int d\vec{r} d\vec{r}\,' \Big(
e\,B(\vec{r})\,\Pi_{\pm} G_{i t}(\vec{r},\vec{r}\,')x' G_{\beta-i t}
(\vec{r}\,',\vec{r})  -  (i t \to i t+\beta) \Big)
\end{equation}
To derive (\ref{time}), the identity
\begin{equation}\label{com}
[H,\Pi_{\pm}]=\mp\left\{eB(\vec{r})\Pi_{\pm} 
+ \big[\Pi_{\pm},\frac{e\,B(\vec{r})}{2}\pm V(\vec{r})\big]\right\}
\end{equation}
has been used, which is valid in general
for an Hamiltonian $H=(\vec{p}-e \vec{A})^2 /2 +V(\vec{r})$. But here, due
to the particular structure of the Hamiltonian (\ref{AC}) where 
$V(r)=-eB(\vec{r})\sigma_z/2$, we have
\bea
[H_u,\Pi_+] &=& -eB(\vec{r}) \, \Pi_+ \\{} % attention a ne pas enlever le {}. Il est utile
[H_d,\Pi_-] &=&  eB(\vec{r}) \, \Pi_- 
\eea
Therefore, (\ref{time}) is valid if, when considering $H_u$, one computes
$\dot\sigma^+$, and, when considering $H_d$, one computes  $\dot\sigma^-$.
In both cases,
we wish to emphasize the appearance of the local magnetic field $B(\vec r)$ 
in (\ref{time}). In the homogeneous field case, it factors
out from the space integrals. In the magnetic impurity case, it is a
sum of $\delta(\vec r-\vec r_i)$ functions, greatly simplifying the space
integrals.

We are looking at the Fourier transform of
$\sigma(t)$ in frequency space,
and more particularly at zero frequency  $\omega=0$ (i.e. the time independent or static response).
The Fourier transform of $\sigma(t)$ is defined by
\begin{equation}
\sigma(\omega)=\lim_{\epsilon\to 0^+} \int_0^{\infty} dt\,\sigma(t)\,
e^{(i \omega - \epsilon)t}
\end{equation}
where $\epsilon$ is a regulator which has to be set to zero at the end.
The Fourier transform of (\ref{time}) rewrites
\begin{eqnarray}\label{fourier}
-(i\omega-\epsilon)\sigma_{\beta}^{\pm}(\omega)&=
&{e^2\over V}\Big\{1  \pm  \frac{1}{Z_{\beta}}
\int_{0}^{\infty} dt\,e^{(i \omega - \epsilon)t}\nonumber\\
& &\hspace{-2.5cm}\int d\vec{r}\,d\vec{r}\,' \Big(
eB(\vec{r})\Pi_{\pm} G_{i t}(\vec{r},\vec{r}\,') x' G_{\beta-i t}(\vec{r}\,',\vec{r})
-\ (i t\ \rightarrow\ i t+\beta) \Big)\Big\}
\end{eqnarray}
and
an analogous formula for the conductivity of a gas of electrons
$\sigma_{E_F}^{\pm}(\omega)$
\begin{eqnarray}\label{ferm}
-(i\omega-\epsilon)\sigma_{E_F}^{\pm}(\omega)
={e^2\over V}\bigg\{N(E_F)   \pm
{1\over 2\pi i}\int_{-\infty}^{\infty}dt'{e^{iE_Ft'}\over
t'-i\eta'}\int_{0}^{\infty} dt\:e^{(i \omega - \epsilon)t}
\int d\vec{r}\,d\vec{r}\,' \nonumber\\
\Big(eB(\vec{r})\Pi_{\pm} G_{i t}(\vec{r},\vec{r}\,')x' 
G_{it'+\epsilon'-i t}(\vec{r}\,',\vec{r})
- (i t\to i t+it'+\epsilon') \Big)\bigg\}
\end{eqnarray}

Up to now, the Hamiltonian (\ref{AC}) has been used. However,
one can simplify the expressions above by noting that a potential vector
$\vec A(\vec r)$, in 2d,  can be always rewritten up to a gauge as
\be 
e A_i(\vec r)=\epsilon_{ij}\partial_j\phi(\vec r)
\ee
with the covariant momentum
\bea
\Pi_+&=&-2i(\partial_{\bar z}+\partial_{\bar z}\phi)\\
\Pi_-&=&-2i(\partial_z- \partial_z\phi)
\eea
Let us redefine the wavefunctions~\cite{nous'} as
\be
\psi(\vec r) = U(\vec r) \tilde\psi(\vec r)
\ee
where the non unitary transformation  $U(\vec r)$ respectively read in the
spin up and down cases 
\bea
\label{1} U_u(\vec r)&=&e^{-\phi(\vec r)}\\
\label{2} U_d(\vec r)&=&e^{+\phi(\vec r)}
\eea
One has
\begin{eqnarray}
\label{A'}
U_u^{-1}\Pi_+ U_u&=&\Pi^0_+ \\
\label{B'} 
U_u^{-1}\Pi_- U_u&=&\Pi^0_-
+4i\partial_z\phi        \\ 
\label{C'} 
U_d^{-1}\Pi_+ U_d&=&\Pi^0_+
-4i \partial_{\bar z}\phi \\
\label{D'}
U_d^{-1}\Pi_- U_d&=&\Pi^0_-
\end{eqnarray}
where
the
potential vector  disappears from the covariant
$U_u^{-1}{\Pi}_+U_u=\Pi^0_+$ and $U_d^{-1}{\Pi}_-U_d=\Pi^0_-$ 
operators, which narrow down to the 
free covariant momentum operators 
$\Pi^{0}_+=-2i\partial_{\bar z}$ and  $\Pi^0_-=-2i\partial_z$.
The  Hamiltonian ${\tilde H}=U^{-1} H U$ acting on $\tilde{\psi}$ rewrites 
\bea
{\tilde H}_u&=&-2\partial_z\partial_{\bar z} +4\partial\phi\:\partial_{\bar z}\\
{\tilde H}_d&=&-2\partial_z\partial_{\bar z} -4{\partial_{\bar z}}
\phi\:\partial_z
\eea
The propagator for the Hamiltonian $H$ being defined in
(\ref{propag}) as
$G_{\beta}(\vec r,\vec r\,')=\bra{\vec r}e^{-\beta H}\ket{\vec r\,'}$, one has
\be 
\tilde{G}_{\beta}(\vec r,\vec r\,')\equiv\bra{\vec r}e^{-\beta \tilde{H}}\ket{\vec r\,'}
={U(\vec r\,')\over U(\vec r)} G_{\beta}(\vec r,\vec r\,')
\ee
It follows that 
the Hamiltonians $H_u$,
$\tilde H_u$, on the one hand, $H_d$,
$\tilde H_d$, on the other hand, are equivalent,
and can be
indifferently used for computing thermodynamical observables, i.e.
traces of product of operators,
as the partition
function, the density of states, or the  conductivity.
The partition function rewrites as
\be\label{finalee} Z_{\beta}=\int d\vec r\:\tilde G(\vec r,\vec r)\ee
and the conductivity becomes
\begin{eqnarray} \label{finale}
\sigma^{\pm}_{\beta}(t)&=&  \frac{e^2}{V}
\theta(t){1\over Z_{\beta}}
\int d\vec{r}\,d\vec{r}\,'
\big\{
U^{-1}(\vec r)\Pi_{\pm} U(\vec r)\tilde {G}_{i t}(\vec{r},\vec{r}\,')
x' \tilde {G}_{\beta-i t}(\vec{r}\,',\vec{r})
\nonumber\\
&& \hspace{5cm}-\ (i t\ \rightarrow\ i t+\beta)\big\}\end{eqnarray}
or
\begin{eqnarray} \label{final}
\dot{\sigma_{\beta}}^{\pm}(t)&=&\frac{e^2}{V}\,\delta(t)  \pm  \frac{e^2}{V}
\theta(t){1\over Z_{\beta}}
\int d\vec{r}\,d\vec{r}\,'
\Big( eB(\vec{r})
U^{-1}(\vec r)\Pi_{\pm} U(\vec r)\tilde {G}_{i t}(\vec{r},\vec{r}\,')
x' \tilde {G}_{\beta-i t}(\vec{r}\,',\vec{r})
\nonumber\\
&& \hspace{5cm}-\ (i t\ \rightarrow\ i t+\beta)\Big)
\end{eqnarray}
It is  remarkable that the covariant momentum operator
$U^{-1}(\vec r)\Pi U(\vec r)$ appearing 
in $\sigma$ or $\dot\sigma$ in
(\ref{finale},\ref{final}) reduces to 
the free momentum provided that one considers $\sigma^+$ 
in the spin up case, and $\sigma^-$ 
in the spin down case. It follows that it is appropriate,
using ${\tilde H}$, to compute
$\sigma^+$ in the spin up case, and $\sigma^-$ in the spin down case.
In the sequel, we will essentially concentrate on the spin down coupling.
However, one should keep in mind that computations for the spin up case
should follow the same lines.

One finally extracts the real part of the transverse and longitudinal 
conductivity
\begin{equation}\label{real}
\re\sigma_{xx}(\omega)\pm i \re\sigma_{yx}(\omega)=\frac{\sigma^{\pm}(\omega)+
\sigma^{\pm}(-\omega)}{2}
\end{equation}
(\ref{real}) is a consequence of (\ref{condcompl}) and is valid both for
the Boltzmann 
(\ref{fourier}) or Fermi (\ref{ferm}) cases.

%%%%%%%%%%%%%%%%%%%%%%%%%%  LANDAU %%%%%%%%%%%%%%%%%%%%%%%%%%%%%%%%%%% 

\section{Hall conductivity for a homogeneous magnetic field}

As a warm up exercise, let us rederive the Hall conductivity
$\sigma_{E_F}(\omega=0)$ in the
homogeneous magnetic field case. By convention, and without any loss of
generality, we assume $eB>0$, and
denote
by $\omega_c=+eB/2$ half the cyclotron frequency.
The Landau Hamiltonian is
\bea 
H_{u}^L={1\over 2}\Pi^L_-\Pi^L_+ =
{1\over 2}\left(\vec p-\omega_c\vec k\times \vec r\right)^2- {\omega_c}\\
H_{d}^L={1\over 2}\Pi^L_+\Pi^L_- =
{1\over 2}\left(\vec p-\omega_c\vec k\times \vec r\right)^2+{\omega_c}
\eea
where
 $\Pi^L_-=-2i(\partial_z-{1\over 2}\omega_c  \bar z)$ is the
 covariant Landau momentum.
 The constant spin couling shift $\mp \omega_c$
 has  clearly no influence on the conductivity.
 The shifted Landau propagator reads ($+$($-$) corresponds to spin $u$ ($d$))
\be\label{proplandau} 
G^L_{\beta}(\vec r,\vec r\,')={\omega_c e^{\pm\beta\omega_c}\over 2\pi\sinh (\beta
\omega_c)}
e^{ -{\omega_c\over 2}(\vert z-z'\vert^2\coth (\beta
\omega_c)+z'\bar z-\bar z' z)}
\ee
Accordingly, the shifted partition function is
\be Z^L_{\beta}=V{\omega_c\over
2\pi}{e^{\pm\beta\omega_c}\over\sinh(\beta\omega_c)}\ee
Its Fermi transform is the Landau density of states, namely in the spin up
case
\be \rho^L(E)={1\over 2\pi }\int_{-\infty}^{\infty}dt'e^{iEt'}Z^L_{it'+\epsilon'}
=V{\omega_c\over \pi}\sum_{n=0}^{\infty}\delta(E-2n\omega_c )\ee
and in the spin down case
\be \rho^L(E)={1\over 2\pi }\int_{-\infty}^{\infty}dt'e^{iEt'}Z^L_{it'+\epsilon'}
=V{\omega_c\over \pi}\sum_{n=0}^{\infty}\delta(E-2(n+1)\omega_c )\ee

As announced above, let us concentrate on the spin down case, i.e. on 
$\sigma^-$ (with a  homogeneous magnetic field, computing 
$\sigma^+$ would lead in the spin up case to  the same
 result).
To compute  the thermal Boltzmann conductivity 
(\ref{thermal}), using 
\be \Pi^L_- G^L_{\beta}(\vec r, \vec r\,')= i\omega_c( \bar z- \bar z')(
\coth (\beta\omega_c) +1  )  G^L_{\beta}(\vec r, \vec r\,')\ee
one has to evaluate  space integrals like
\be\label{ill}
{1\over V}\int d\vec{r}\, d\vec{r}\,'
(\bar z- \bar z') G^L_{it}(\vec r,\vec r\,')x' G^L_{\beta-i t}(\vec r\,',\vec r)
\ee
Since, in the Landau case, $G^L_{it}(\vec r,\vec r\,')G^L_{\beta-i t}(\vec r\,',\vec r) $ is a 
function of $|z-z'|$ only, (\ref{ill}) might seem naively ill defined
(infinite by power counting,
vanishing by symmetry). Still, one can proceed if, looking at
(\ref{local}), one pays attention to first perform the $\vec r\,'$
integral, leading to the local conductivity $\sigma^L_{\beta}(\vec r,t)$, which 
upon integration on $\vec r$,  yields  the global conductivity.
In the homogeneous magnetic field case, 
the $\vec r\,'$ integration in (\ref{ill}) gives, as it should, 
a result which is independent of $\vec r$, the manifestation of the translation
invariance of the system.
The remaining $\vec r$ integral simply factorizes a volume factor, which
gives an unambiguous meaning to the divergence in (\ref{ill}). One finds
\be\label{Landres} 
{\sigma^L_{\beta}}^-(t)
=\theta(t){e^2\over V}e^{2i\omega_c t}
\ee

We would like to emphasize that (\ref{Landres})  can be trivially 
recovered if one now looks at $\dot
\sigma^{L-}_{\beta}(t)$, instead of $\sigma^{L-}_{\beta}(t)$.
Indeed, in (\ref{time}) $B$ factors out from the space integrals. So,
(\ref{time} rewrites as
\be 
\dot\sigma^{L-}_{\beta}(t)={e^2\over V}\delta(t) +2i\omega_c 
\sigma^{L-}_{\beta}(t)
\ee
which can be directly integrated to (\ref{Landres}), without any need of properly
defining, in the thermodynamic limit, divergent space integrals.

In principle, one  proceeds by taking the Fourier and Fermi transforms of 
(\ref{Landres}). 
 The former reads
\be \label{tf}
\sigma^{L-}_{\beta}(\omega)={e^2\over V}{1\over
\epsilon-i(\omega+2\omega_c)}=
{e^2\over V}\left\{{\cal PP}{i\over
\omega+2\omega_c}+\pi\delta(\omega+2\omega_c)\right\}\ee
and the latter rewrites
\be 
\sigma^{L-}_{E_F}(t)=\theta(t){e^2\over V}e^{2i\omega_ct}{1\over 2\pi i}
\int_{-\infty}^{\infty}dt'{e^{iE_Ft'}\over t'-i\eta'} Z_{it'+\epsilon'}
\ee
which yields the infinite series $\sum_n \theta(E_F-2\omega_c(n+1))$. But this series 
is meaningless, because $E_F$ is not defined due to the zero
width of the Landau levels.

However, coming back to (\ref{Landres}) or (\ref{tf}), one sees that the thermal conductivity 
does not depend on $\beta$, implying that the conductivity is the same
for each Landau state. It follows that, 
for a gas of electrons at zero temperature, the
conductivity is  obtained by multiplying (\ref{Landres}) or (\ref{tf}) by $
N(E_F)$.
 From (\ref{real}), one infers the transverse conductivity 
\be\label{landaures}
{\re\sigma^L_{E_F}(\omega)|}_{yx}=-N(E_F)\frac{e^2}{V}
\frac{2\omega_c}{4\omega_c^2-\omega^2} 
\ee
In (\ref{landaures}), the limit $\omega_c\to 0$  is properly defined only if
one keeps $\omega\ne 0$, in which case it
vanishes, as it should.
The Hall conductivity finally reads
\be
\re\sigma^L_{E_F}(\omega=0)|_{yx}=-N(E_F)\frac{e}{ V}{1\over B}
\ee
This is  the classical straight line, showing
no plateaus in the Hall conductivity as a function of the number of
electrons $N(E_F)$, or of
the inverse magnetic field $1/B$. 

The longitudinal conductivity 
  is
\be\label{sans}
\re\sigma^L_{E_F}(\omega)|_{xx}=N(E_F)\frac{e^2}{ V}{\pi\over
2}\left\{\delta(\omega+2\omega_c)+\delta(\omega-2\omega_c)\right\}
\ee

%%%%%%%%%%%%%%% ONE VORTEX   &    VORTEX + B  %%%%%%%%%%%%%%%%%%%%%%%%%%%%

\section{Hall conductivity for one vortex and for one vortex + a homogeneous
$B$ field}

On the one hand, these two non trivial examples of magnetic systems are
interesting because their conductivities can be
entirely computed, thanks to the local magnetic field $\delta(\vec r)$
functions in the space integrals of $\dot \sigma_{\beta}(t)$. On the
other hand, 
they  are respectively related, in the random magnetic impurity problem, to the
  conductivity of the mean magnetic field, 
  and to  perturbative corrections at first order in $\alpha$
to the conductivity of the mean magnetic field. 

As in the Landau case, we set $eB>0$,  thus $\omega_c=+eB/2$.
We consider a vortex at the origin with flux $\phi$. The
coupling of the electron to the vortex is $e\phi/2\pi=\alpha$.
This system is periodic in $\alpha$ of
period 1, therefore $\alpha$ can always be choosen
in the interval $[-1/2,1/2]$.  If $B=0$,
the interval of definition of $\alpha$ can be restricted to $[0,1/2]$, since
reversing the sign of the flux does not change the system.
If $B\ne 0$,
 $\phi$ can be either parallel ($\alpha>0$), or antiparallel ($\alpha<0$) to the
magnetic  field, which
imposes an orientation to the plane. Clearly, 
the physics depends on the sign of $\alpha$. We will 
focus on the physical situation where the flux and the
magnetic
field are parallel
$\alpha\in [0,1/2]$. Indeed,
this is what happens in the random magnetic impurity problem,
where the mean magnetic field $\mean{B}$ is  built by the vortices $\phi$
carried by the impurities.

In the symmetric gauge, the Hamiltonian $H_d$ reads
\begin{equation}\label{vbd}
H_{d}=\frac{1}{2}\left(\vec{p}-\omega_c \vec k\times\vec r-
\alpha \frac{\vec{k}\times\vec{r}}{r^2}\right)^2
+(\pi\alpha\delta(\vec{r})+\omega_c)
\end{equation}
where $\vec k$ is the unit vector
perpendicular to the plane. 

The coupling of the electron to the
infinite magnetic field inside the vortex defines in a non
ambiguous way the short distance behaviour of the electron wavefunctions
at the origin. This need for a proper 
characterisation~\cite{lozano}~\cite{nous'} of boundary conditions
at the location of the vortex is related to the fact 
that perturbation theory in $\alpha$ is ill defined for the
Hamiltonian (\ref{vbd}) in the zero orbital
momentum $m=0$ sector. In fact,  wavefunctions of zero orbital
momentum $m=0$ are  affected by the short distance regularisation
$\pi\alpha\delta(\vec r)$. In the case of the spin down Hamiltonian $H_d$,
the contact term is repulsive,
indicating that the wavefunctions have to vanish when $r\to 0$ as
quickly as $r^{\alpha}$. It follows that in the $m=0$ sector,  only
 regular wavefunctions at the origin should be retained, as in 
 the standard Aharonov-Bohm prescription~\cite{AB}. 
 It means that the electron cannot penetrate inside the vortex,
a quite reasonable physical  situation.

\subsection{Partition function and density of states}

The free and  Landau partition functions diverge  like the
volume  in the
thermodynamic limit (continuous spectra), and corrections
 due to the vortex are  subleading in
volume. To give an unambiguous meaning to these corrections, one needs
to regularize infrared divergences. This can be achieved by adding to the
Hamiltonian an
harmonic regulator, ${1\over 2}\omega_o^2 r^2$, which has to be set to zero
$\omega_o\to 0$ at
the end. 
The Hamiltonian (\ref{vbd}) with the additional harmonic term has
the spectrum 
\begin{equation}m\ne 0\quad\
E_{n,m}=(2n + |m-\alpha|+1)\omega_t - (m-\alpha)\omega_c+\omega_c
\end{equation}
\begin{equation}m= 0\quad\
E_{n,0}=(2n +\alpha+1)\omega_t +\alpha\omega_c+\omega_c
\end{equation}
where $\omega_t^2=\omega_c^2+\omega_o^2$.
The partition function 
rewrites
\begin{equation}
Z^{\omega_o}_{\beta}(B,\alpha)
-Z^{\omega_o}_{\beta}(B,0)={e^{-\beta\omega_c}\over
2\sinh(\beta\omega_t)}
\left\{
  \frac{e^{-\alpha\beta(\omega_t+\omega_c)}-1}
       {1-e^{-\beta(\omega_t+\omega_c)}}
  -\frac{e^{\alpha\beta(\omega_t-\omega_c)}-1}
           {1-e^{\beta(\omega_t-\omega_c)}}
\right\}
\end{equation}
The thermodynamic limit,
 $\omega_o\to 0$,  i.e.
$\omega_t-\omega_c\to 0$, yields
\be \label{ZBVd}
Z_{\beta}(B,\alpha)-Z_{\beta}(B,0)=
{e^{-\beta\omega_c}\over 2\sinh(\beta\omega_c)}
\left\{
  \alpha+{e^{-2\alpha\beta\omega_c}-1\over 1-e^{-2\beta\omega_c}}
  \right\}
\ee
The Fermi transform  of (\ref{ZBVd}) gives  the density of states 
\begin{eqnarray}\label{mis}
\rho(E,B,\alpha)-\rho(E,B,0)=
&-&\sum_{n=0}^{\infty}(n+1-\alpha)\,\delta(E-2(n+1)\omega_c)\nonumber\\
&+&\sum_{n=0}^{\infty}
(n+1)\,\delta(E-2(n+1+\alpha)\omega_c)\end{eqnarray}
In the limit $B\to0$, (\ref{ZBVd}) becomes
\be\label{depletion'}
Z_{\beta}(0,\alpha)-Z_{\beta}(0,0)={\alpha(\alpha-1)\over 2}
\ee
and accordingly
\be\label{mis'}
\rho(E,0,\alpha)-\rho(E,0,0)=\frac{\alpha(\alpha-1)}{2}\delta(E)
\ee
i.e. the usual Aharonov-Bohm depletion of  states~\cite{CMO} at the bottom of the
spectrum with respect to the free density of states $\rho(E,0,0)\equiv 
\rho^0(E)$.

The physical meaning of (\ref{mis}) is that on each Landau level $E_n=2(n+1)\omega_c$,
$n+1-\alpha$ states states disappear and $n+1$
appear at energy $2(n+\alpha+1)\omega_c$.
When $B=0$, the sole effect of the vortex in (\ref{mis'}) is that 
$\alpha(1-\alpha)/2$ states have left the bottom of the
spectrum.

\subsection{Hall conductivity for one vortex}
Let us first concentrate on the vortex alone.
The propagator of (\ref{vbd}) with $B=0$ reduces to the  
standard Aharonov-Bohm propagator
\begin{equation}
G_{\beta}(\vec{r},\vec{r}\,')=\frac{1}{2\pi \beta}
e^{-\frac{1}{2\beta}(r^2+{r'}^2)}
\sum_{m=-\infty}^{+\infty} I_{\vert m-\alpha
\vert}\left(\frac{rr'}{\beta}\right) e^{i m (\theta-\theta')}
\end{equation}
where the
$I_{\nu}(z)$'s are the modified Bessel functions.

One expects that the Hall conductivity is antisymmetric with respect to
$\alpha\to 1-\alpha$, because of the antisymmetry $\alpha\to -\alpha$ (the
transverse
conductivity has to change its sign when the direction of the local
magnetic field is reversed) and
the periodicity. In particular it has to  vanish when
$\alpha=1/2$.
One has to compute (\ref{time}), where
$eB(\vec{r})=2\pi\alpha\,\delta(\vec{r})$ and
$\Pi_-=-i (2 \partial_z-\alpha / z)$.
The $\delta(\vec{r})$ appearing in
 $\dot \sigma_{\beta}^-(t)$  greatly simplifies a computation
 otherwise untractable. Indeed, only the $m=0$ and
$m=1$ orbital terms in the propagators eventually survive to yield
\begin{equation}\label{result}
\dot\sigma_{\beta}^-(t)=\frac{e^2}{V}\delta(t)+\frac{e^2}{V}\theta(t)
\frac{1}{\beta^2 Z_{\beta}}
{e^{i\pi\alpha}}{\sin(\pi\alpha)\over \pi}( 
t^\alpha\,(t+i\beta)^{1-\alpha}-
t^{1-\alpha}\,(t-i\beta)^\alpha)
\end{equation}
Its Fourier transform reads
\bea
\sigma_\beta^-(\omega)={e^2\over V }\frac{i}{(\omega+i\epsilon)}
\bigg\{1-{1\over Z_{\beta}} 
e^{i\pi\alpha}{\sin(\pi\alpha)\over \pi}\Big(\Gamma(1+\alpha) \Psi(1+\alpha,3;\beta(\omega+i\epsilon))\nonumber\\ 
     -\Gamma(2-\alpha) \Psi(2-\alpha,3;-\beta(\omega+i\epsilon))
\Big)\bigg\}
\eea
where the $\Psi(a,b;z)$'s are  the unregular confluent hypergeometric functions.
To get the Hall conductivity $\re\sigma(\omega)|_{yx}$ in the
$\omega\to 0$ limit (static response), special attention must
be paid to the $\epsilon\to 0^+$ limit,  because of the logarithm
that appears in the low $\beta\omega$ expansion
of $\Psi(a,b;\beta(\omega+i\epsilon))$. One obtains (the electron
mass, $m_e$, and $\hbar$ have been reintroduced)
\be\label{vortexres}
\re\sigma_{\beta}(\omega)|_{yx}= \frac{\hbar e^2}{m_e^2 V^2}
\frac{\sin(2\pi\alpha)}{\omega^2}
\Big(
1+\frac{\alpha(1-\alpha)}{2}
(\beta\omega)^2\ln(\beta\omega)
+\cdots \Big)\ee
Thus, a
single vortex in the plane is sufficient to 
display a Hall conductivity, an explicit
$P$ violating effect (see~\cite{deja} and references therein for early
tentatives on the subject). Of course, it
is subleading in volume
$\sigma\simeq  1/V^2$
when compared to the homogeneous
$B$-field case $\sigma\simeq 1/V$. However, when, in the random magnetic
impurity problem,
comparaison will be made with the mean magnetic
field, the $1/V^2$ factor  will  be multiplied by the number
of impurities $N$, 
yielding, in the thermodynamic limit, an adequate  $\rho/V$ behavior.
Also worth mentionning is the
leading $1/\omega^2$ divergent behaviour of (\ref{vortexres}) in the  $\omega
\to 0$ limit,
which will also be  related to the dynamical conductivity of the mean magnetic field.

For a gas of electrons coupled to the vortex at zero temperature, one has to
compute
\begin{eqnarray}\label{fer}
-(i\omega-\epsilon)\sigma_{E_F}^{-}(\omega)
={e^2\over V}\bigg\{N(E_F)+
{1\over 2\pi i}\int_{-\infty}^{\infty}dt'{e^{iE_Ft'}\over
t'-i\eta'}\int_{0}^{\infty} dt e^{(i \omega - \epsilon)t}
\frac{1}{(it'+\epsilon')^2}\nonumber\\
\frac{e^{i\pi\alpha}}{\Gamma(\alpha)\Gamma(1-\alpha)}\big( 
t^\alpha\,(t-t'+i\epsilon')^{1-\alpha}-
t^{1-\alpha}\,(t+t'-i\epsilon')^\alpha\big) \bigg\}
\end{eqnarray}
At small frequency,
one should look at the leading high $t$ behaviour of (\ref{result})
in which $\beta$ has been  replaced by $it'+\epsilon'$. Since  
$t^\alpha\,(t-t'+i\epsilon')^{1-\alpha}-
t^{1-\alpha}\,(t+t'-i\epsilon')^\alpha\simeq -t'$, one finds 
a conductivity proportional to $N(E_F)$
\be \sigma_{E_F}(\omega\to 0)\simeq N(E_F){e^2\over V}\left\{ {1\over
\epsilon-i\omega}+{1\over V}2ie^{i\pi\alpha}\sin (\pi\alpha){1\over
(\epsilon-i\omega)^2}\right\}\ee
Thus, in the limit $\omega\ll E_F$, 
\be \label{oui}\re\sigma_{E_F}(\omega)|_{yx}=N(E_F){e^2\over V^2}\sin(2\pi\alpha){1\over
\omega^2}\ee
and
\be \label{non}\re\sigma_{E_F}(\omega)|_{xx}=N(E_F) {e^2\over V^2} 2\sin^2(\pi\alpha) {1\over
\omega^2}\ee

\subsection{Hall conductivity for one vortex + a homogeneous magnetic field} 

The propagator of the
Hamiltonian (\ref{vbd}) (spin down, $\alpha \in [0,1/2]$) is
\begin{equation}
G_{\beta}(\vec{r},\vec{r}\,')=\frac{\omega_c e^{-\beta\omega_c}}{2\pi \sinh \beta\omega_c}
e^{ -\frac{\omega_c}{2} \coth \beta\omega_c (r^2+{r'}^2) }
\sum_{m=-\infty}^{+\infty} I_{\vert m-\alpha
\vert}\left(\frac{\omega_c\,r\,r'}{\sinh \beta\omega_c}\right) 
e^{\beta\omega_c(m-\alpha)}e^{i m (\theta -\theta')} 
\end{equation}
As in the one vortex case, and exactly for the same reason, only terms of angular momentum
$m=0$ and $m=1$ survive in $\dot\sigma^-_{\beta}(t)$
\be\label{B+vortres}
\dot\sigma_{\beta}^-(t)
=\frac{e^2}{V}\delta(t)+2i\omega_c\sigma_\beta^-(t)
+{e^2\over V}i\theta(t)
\frac{\omega_c e^{-\beta\omega_c(\alpha+1)}}{Z_{\beta}\sinh^2(\beta\omega_c)}
{\sin(\pi\alpha)\over \pi}G(t)
\ee
where
\be
G(t)=\theta(t)\left\{e^{it\omega_c}\left(\sinh it\omega_c\right)^\alpha \left(\sinh
(\beta-it)\omega_c\right)^{1-\alpha} - (it\to it+\beta)\right\}
\ee
One can check that  the limit $\omega_c\to 0$ of $\dot\sigma_{\beta}^-(t)$ 
in (\ref{B+vortres}) gives
(\ref{result}), as it should. 
Its Fourier transform is
\be\label{soBV}
\sigma^-_{\beta}(\omega)={1\over \epsilon-i(\omega+2\omega_c)}
\frac{e^2}{V}\bigg\{
1+i{1\over Z_{\beta}}\frac{\omega_c e^{-\beta\omega_c(\alpha+1)}}{\sinh^2(\beta\omega_c)}
{\sin(\pi\alpha)\over \pi}G(\omega)\bigg\}
\ee
where $G(\omega)$ is the Fourier transform of $G(t)$.
The
 thermal  conductivity,
splits in two terms,
the usual Landau term for the magnetic field,  and a
term due to the
vortex alone. Since
$G(t)$ is periodic,  one expects the conductivity to be regular at low
frequency. One has 
\be
G(\omega)=\frac{1}{1-e^{i2\pi\omega/\omega_c}}\int_0^{2\pi/\omega_c}dt\,
G(t)e^{i\omega t}
\ee
Since $G(t)^*=G(-t)$, $G(\omega)$ is purely imaginary. It follows that,
in (\ref{soBV}), no contribution to the real part of the longitudinal conductivity 
has to be expected from the vortex.
Furthermore, $G(\omega)$ is finite when $\omega\to 0$, since $\int_0
^{2\pi/\omega_c} dt\, G(t)=0$.
These two points are to be contrasted with the pure vortex case where the
longitudinal and transverse conductivities
have been shown to diverge at low frequency (\ref{oui},\ref{non}). It means that, as in
the pure Landau case, the limit $\omega_c\to 0$
is not smooth when $\omega\to 0$. In other words, if one wants  to
recover (\ref{oui},\ref{non}) from the present analysis, one should, while keeping
$\omega\ne 0$,
take the limit $\omega_c\to 0$, and then consider the small $\omega$ expansion.
Here, we are considering the other situation where $\omega\to 0$, while
$\omega_c$ is kept finite.

One is interested by the conductivity for a gas of electrons,
i.e. the Fermi transform of (\ref{soBV}), where $\beta$ has to be
replaced by a complex time $\beta \to i t'+\epsilon'$.
It happens that the  small  $\alpha$ expansion
of $G(\omega=0)$ when $\beta$ is imaginary has no term in $\alpha$
\be G(\omega=0)={i}{1\over 2\omega_c}\sinh(i\omega_c t')+ O(\alpha^2)+\ldots \ee
Note, as a remark,  that, when $\alpha=0, G(\omega)={i}{1\over
2\omega_c+\omega}\sinh(i\omega_c t')$.

It follows that the small $\alpha$ expansion of the Fermi transform of the
static thermal conductivity (\ref{soBV}) reads
\be\label{BVcond}
\sigma^-_{E_F}(\omega=0)=\int_{-\infty}^{+\infty}\frac{dt'}{2i\pi}\frac{e^{iE_Ft'}}{t'-i\eta'}
\frac{i e^2}{2V\omega_c}\left\{
  Z_{it'}+{Z_{it'}^L\over V}\left(
    -\frac{\pi\alpha}{\omega_c}  +\pi\alpha^2{it'}+\cdots \right)
  \right\}\ee
  where the partition function $Z_{it'}$ given in (\ref{ZBVd}) should also be understood as
  expanded at first order in $\alpha$.
  It is not surprising that
 in  (\ref{BVcond}), the first order correction to the conductivity
appears  as a mean magnetic field contribution
of the vortex  ${\pi\alpha}/{V}$ to the magnetic field $\omega_c$
\be
\sigma^-_{E_F}(\omega=0)=N(E_F)\frac{ie^2}{2V}(
  \frac{1}{\omega_c}-\frac{\pi\alpha}{V}
  \frac{1}{\omega_c^2})+\cdots 
= N(E_F)\frac{ie^2}{2V}\frac{1}{\omega_c+{\pi\alpha}/{V}}+\cdots 
\ee
to be compared to the Landau conductivity (\ref{tf}).

\section{Hall conductivity for the magnetic impurity problem }

\subsection{The magnetic impurity problem}
 
Let us now onsider an electron coupled to 
the vector potential of a random distribution $\{\vec r_i, i=1,2,....,N\}$
of $N$ vortices at position $\vec r_i$, carrying a flux $\phi$, 
hereafter called magnetic impurity.  
The thermodynamic limit is $N\to \infty, V\to\infty$, with a finite 
impurity density $\rho=N/V$.
The
distribution is Poissonian, i.e. 
point magnetic impurities dropped on the plane randomly without correlation. 
It means that the infinitesimal
probability $dP(\vec r_i)$ of finding an impurity at position $\vec r_i$ is 
\be\label{poisson} dP(\vec r_i)={d\vec r_i\over V}\ee

One would like to see in which way this system could be compared
with the naive mean field approximation where the local magnetic field
$eB(\vec r)=2\pi\alpha\sum_{i=1}^N\delta(\vec r-\vec r_i)$ is replaced by its
mean value \footnote{
If one takes a 2 dimensional
impurity density of the order of the density of current carriers 
$\rho=4. 10^{15}m^{-2}$, one obtains, for $\alpha=1/2$,
a  magnetic field   precisely in the
experimental range of the Quantum Hall Effect
\be \mean{B}={h\rho\alpha\over e}\simeq 10T\ee
}
\be\label{??} \mean{ B} V=N\phi \quad {\rm i.e.\quad in\quad the\quad thermodynamic
\quad limit} \quad e\mean {B}=2\pi\rho\alpha\ee
Two approaches have been  used~\cite{nous} to study this problem: a path integral Brownian motion 
approach
 (or its discretised version, random walk in the plane, which allows
 for numerical simulations), which focuses on winding properties of Brownian
 paths in the plane, and a quantum mechanical Hamiltonian approach, 
which relies mainly on perturbative expansions of the average partition
function (small $\alpha$, high
temperature, etc). 
In the sequel, we will use the quantum mechanical Hamiltonian approach, that we summarize now.

 In the symmetric gauge,
 the Hamiltonian for $N$ impurities is
\be\label{11} 
H_d={1\over 2}
\left(\vec p - \sum_{i=1}^N\alpha{\vec k\times(\vec r-\vec r_i)
\over (\vec r -\vec r_i)^2}
\right)^2
+\pi\alpha\sum_{i=1}^N \delta(\vec r-\vec r_i)
\ee
where $\alpha\in[0,1/2]$, without any loss of generality. 
The need for a spin coupling term has already been discussed
above. Recall that, physically, it corresponds to hard-core impurities
where the electron is
unable to penetrate.

Consider the nonunitary wavefunction redefinitions (\ref{2})
\be\label{new1}
\psi(\vec r)=\prod_{i=1}^N\vert\vec r-\vec r_i\vert^{\alpha}
\tilde{\psi}(\vec r)\equiv U_d(\vec r)\psi(\vec r)
\ee
(\ref{C'},\ref{D'}) become
\bea
\label{C} 
U_d^{-1}\Pi_+ U_d&=&\Pi^0_+ - 2i\alpha\Omega \\
\label{D}
U_d^{-1}\Pi_- U_d&=&\Pi^0_-
\eea
where the impurity vector potential
$\Omega=\sum_{i=1}^N{1/( \bar z-\bar z_i)}$.
The Hamiltonian $\tilde{H}$ acting on 
$\tilde{\psi}(\vec r)$ rewrites
\bea \label{101}
\tilde{H}_d&=&-2\partial_{\bar z}\partial_z 
-2\alpha\sum_{i=1}^N {1\over \bar z-\bar z_i}\partial_z\\
&=&{1\over 2}\Pi^{0}_+\Pi^{0}_- 
- i\alpha \Omega\Pi^{0}_-
\eea
One can go a bit further by extracting~\cite{nous} the mean value
of $\prod_{i=1}^N\vert\vec r-\vec r_i\vert^{\alpha}$ which precisely
yields,
in the thermodynamic limit, the  mean magnetic field Landau
exponential factor $e^{{1\over2} \mean{\omega_c} r^2}$
with $\mean{\omega_c}=e\mean{B}/2=\pi\rho\alpha$.
Remember that $\alpha\in[0,1/2]$, so that  $e\mean{B}$, 
which has the sign of $\alpha$, is indeed positive. So let us redefine
\be \label{neww}
\psi(\vec r)=e^{-{1\over2} \mean{\omega_c} r^2}
\prod_{i=1}^N\vert\vec r-\vec r_i\vert^{\alpha}
\tilde{\psi '}(\vec r)\equiv U'_{d}(\vec r)\tilde{\psi '}(\vec r)
\ee
and obtain
\bea
(U'_d)^{-1}\Pi_+ U'_d&=&\Pi^{<L>}_+
-2i\alpha(\Omega-\langle\Omega\rangle) \\
(U'_d)^{-1}\Pi_- U'_d&=&\Pi^{<L>}_-
\eea
Thus, 
\be\label{extract} 
\tilde{H'}_d
={1\over 2}\Pi^{<L>}_+\Pi^{<L>}_- 
- i\alpha (\Omega-\mean{\Omega})\Pi^{<L>}_-
\ee
where $\Pi^{<L>}_-=-2i(\partial_z-\mean{\omega_c}\bar z/2)$ is the covariant momentum operator in the
 symmetric gauge for the Landau Hamiltonian of the mean magnetic field, and
$\mean{\Omega}=\pi\rho z$  is the mean value of $\Omega$.
 Therefore,
 (\ref{extract})
is the Landau Hamiltonian for the mean field  $\mean{B}$  up to
a $\mean{\omega_c}$ shift, encoding
the mean magnetic field contributions, plus a correction
due to disorder, gathered in
the interaction  term  ${\tilde{V}}'=-i\alpha(\Omega-\mean{\Omega})\Pi^{<L>}_-$.

The Hamiltonians $H_d$,
$\tilde H_d$ and   $\tilde {H'}_d$  are equivalent,
and can be
indifferently used for computing thermodynamical observables as the average partition
function, the average density of states, or the average conductivity.
However, quadratic
interactions in the vector potential have disappeared from the Hamiltonians
(\ref{101},\ref{extract}), 
making easier the average over disorder.
In the case of the  Poisson probability distribution 
(\ref{poisson}), the average  can be done using identities like
\begin{eqnarray}\label{meanrules} 
\int d\bar z_idz_i {1\over \bar z-\bar z_i}&=&\pi z\\
\int d\bar z_idz_i{1\over \bar z-\bar z_i}
{1\over \bar z'-\bar z_i}&=&\pi ({z\over \bar z'-\bar z}
+{z'\over \bar z-\bar z'})\quad\\
&\vdots&\nonumber
\end{eqnarray}

What general information can be extracted from (\ref{11},\ref{extract})
after averaging over the Poissonian disorder?

i) because of the periodicity $\alpha\to \alpha+1$, and, in the absence of any
orientation to the plane, because of the symmetry
$\alpha\to-\alpha$, observables such as the partition function or the
density of states, should be a function of $\alpha(1-\alpha)$ only. The
transverse conductivity, on the other hand, because of the electric field, should change
its sign when $\alpha\to-\alpha$, implying that it vanishes
when $\alpha=1/2$. 

ii) any observable can be expanded in power series in $\alpha$ and $\rho$
as $\sum_{n,m}\rho^n\alpha^m, m\ge n$. The constraint $m\ge n$ stems from
the fact the number of vertex is necessarily larger than the number of
impurities (an electron can interact at different times with the same
impurity). This series can always be rewritten as
$ \sum_{p=0}^{\infty}\alpha^p f_p(\alpha\rho)$

iii) the $ \rho\alpha^n$ and $(\rho\alpha)^n$ terms are exactly known:
the former correspond to the entirely solvable one impurity problem
(standard Aharonov-Bohm problem), whereas the latter is obtained, 
if, in the Hamiltonian ${\tilde H'}_d$, one sets $\Omega=\mean{\Omega}$,
thus reducing the problem to the average Landau problem.
In the absence of disorder,
any observable is necessarily a power
series in
leading order $(\rho\alpha)^n$ of the mean magnetic field. 

iv) corrections coming from disorder are  necessarily of the
type $\rho^n\alpha^m, m>n$.
To have
information on these corrections, perturbation theory is  needed.
However, if perturbation theory is meaningful for  the Hamiltonian
(\ref{11}),  still ultraviolet divergences occur in the perturbative
computation of the
spectrum, which are
shown to cancel only if proper regulators are used~\cite{lozano}~\cite{nous'}.
On the contrary,
the Hamiltonians $\tilde{H}_d$ or $\tilde{H}'_d$,
(\ref{101},\ref{extract}), yield a  perturbation theory  which is  finite
and easier to
handle than whose of (\ref{11}).  If one 
concentrates on the perturbative expansion for the Hamiltonian
$\tilde{H}'_d$, the unperturbed Hamiltonian is the Landau
Hamiltonian for the mean magnetic field, that is to say
$\mean{\omega_c}=\pi\rho\alpha$ is fixed.  It follows that  $\rho^n\alpha^m, m>n$
terms correspond to $\alpha^{m-n}$ corrections to the mean
field. For instance, $\rho\alpha^2$ is actually of order $\alpha$, meaning
that a second order $\alpha^2$ computation
produces a first order $\alpha$ correction.
Conversely, the $\alpha$ correction to the mean field will involve all the 
 $\rho^n\alpha^{n+1}$ terms.

To develop a perturbative expansion for the partition function (\ref{finalee})
or for the
conductivity (\ref{finale},\ref{final}), all what is needed is the perturbative expansion of
the propagator $\tilde{G}'_\beta(\vec r,\vec r\,')=\bra{\vec r}
e^{-\beta\tilde{H}'_d}\ket{\vec r\,'}$ of the Hamiltonian (\ref{extract}).
At order $n$, one has
\bea\label{pertexp}
\delta\tilde{G}'^{(n)}_\beta (\vec r,\vec r\,')=
(-1)^n \int_0^\beta d\beta_1 \int_0^{\beta_1} d\beta_2\cdots 
\int_0^{\beta_{n-1}} d\beta_n \int d\vec r_1\cdots \int d\vec r_n \nonumber\\
G^{<L>}_{\beta-\beta_1}(\vec r,\vec r_1) 
{\tilde V}^{'}(\vec r_1) G^{<L>}_{\beta_1-\beta_2}(\vec r_1,\vec r_2) 
\cdots
{\tilde V}^{'}(\vec r_n) G^{<L>}_{\beta_n}(\vec r_n,\vec r\,')
\eea
where ${\tilde V}^{'}=-i\alpha(\Omega-\mean{\Omega})\Pi^{<L>}_-$ and
$G^{<L>}_{\beta}(\vec r,\vec r\,')$ is the propagator of the mean
magnetic field  (\ref{proplandau}) for the spin down case with
$\omega_c\to \mean {\omega_c}$.
The  only virtue of  $\mean{\Omega}$ in ${\tilde{V}}^{'}$ is to cancel the mean field
$(\rho\alpha)^n$ corrections coming from $\Omega$ alone. This is why one can
ignore $\mean{\Omega}$ if one pays attention to keep only the non mean field
contributions due to $\Omega$. In term of Feynman diagrams, it
means that after averaging over the positions of impurities, one should only
retain the diagrams without isolated impurity legs. It follows that the  correction in $\alpha$
to the mean field expansion  are obtained from the second
order  $\rho\alpha^2\simeq \mean{\omega_c}\alpha$
corrections, with   diagrams involving only one impurity.

Let us illustrate these  considerations
by computing the corrections to the average Landau partition function.
One has to evaluate perturbatively the average second order correction to the
average Landau propagator.
(\ref{pertexp}) yields the correction 
 \bea
\delta\tilde{G}'^{(2)}_\beta(\vec r,\vec r\,')&=&(i\alpha)^2
\int_0^\beta d\beta_1 \int_0^{\beta_1}d \beta_2 \int d\vec r_1 d\vec r_2
G^{<L>}_{\beta-\beta_1}(\vec r,\vec r_1)
\nonumber\\
& &\hspace{-1cm}\sum_{i=1}^N\frac{1}{\bar z_1-\bar z_i} \Pi^{<L>}_{ -} 
G^{<L>}_{\beta_1-\beta_2}(\vec r_1,\vec r_2)
\sum_{j=1}^N\frac{1}{\bar z_2-\bar z_j} \Pi^{<L>}_{ -} 
G^{<L>}_{\beta_2}(\vec r_2,\vec r\,')
\nonumber\\
&=&\begin{picture}(0,0)%
\epsfig{file=pertG2.pstex}%
\end{picture}%
\setlength{\unitlength}{0.00087500in}%
\begingroup\makeatletter\ifx\SetFigFont\undefined
% extract first six characters in \fmtname
\def\x#1#2#3#4#5#6#7\relax{\def\x{#1#2#3#4#5#6}}%
\expandafter\x\fmtname xxxxxx\relax \def\y{splain}%
\ifx\x\y   % LaTeX or SliTeX?
\gdef\SetFigFont#1#2#3{%
  \ifnum #1<17\tiny\else \ifnum #1<20\small\else
  \ifnum #1<24\normalsize\else \ifnum #1<29\large\else
  \ifnum #1<34\Large\else \ifnum #1<41\LARGE\else
     \huge\fi\fi\fi\fi\fi\fi
  \csname #3\endcsname}%
\else
\gdef\SetFigFont#1#2#3{\begingroup
  \count@#1\relax \ifnum 25<\count@\count@25\fi
  \def\x{\endgroup\@setsize\SetFigFont{#2pt}}%
  \expandafter\x
    \csname \romannumeral\the\count@ pt\expandafter\endcsname
    \csname @\romannumeral\the\count@ pt\endcsname
  \csname #3\endcsname}%
\fi
\fi\endgroup
\begin{picture}(2115,1167)(226,-712)
\put(2341,-556){\makebox(0,0)[lb]{\smash{\SetFigFont{12}{14.4}{rm}$\vec r$}}}
\put(1936,-691){\makebox(0,0)[lb]{\smash{\SetFigFont{8}{9.6}{rm}$\beta-\beta_1$}}}
\put(226,-556){\makebox(0,0)[lb]{\smash{\SetFigFont{12}{14.4}{rm}$\vec r\,'$}}}
\put(856,299){\makebox(0,0)[lb]{\smash{\SetFigFont{12}{14.4}{rm}$\vec r_j$}}}
\put(1756,299){\makebox(0,0)[lb]{\smash{\SetFigFont{12}{14.4}{rm}$\vec r_i$}}}
\put(541,-691){\makebox(0,0)[lb]{\smash{\SetFigFont{8}{9.6}{rm}$\beta_2$}}}
\put(1126,-691){\makebox(0,0)[lb]{\smash{\SetFigFont{8}{9.6}{rm}$\beta_1-\beta_2$}}}
\put(721,-421){\makebox(0,0)[lb]{\smash{\SetFigFont{10}{12.0}{rm}$\alpha$}}}
\put(1621,-421){\makebox(0,0)[lb]{\smash{\SetFigFont{10}{12.0}{rm}$\alpha$}}}
\put(811,-646){\makebox(0,0)[lb]{\smash{\SetFigFont{10}{12.0}{rm}$\vec r_2$}}}
\put(1711,-646){\makebox(0,0)[lb]{\smash{\SetFigFont{10}{12.0}{rm}$\vec r_1$}}}
\end{picture}

\eea
The average over impurity positions $\vec r_i$ and $\vec r_j$,
using (\ref{meanrules}), produces either a term
containing  interaction with two
different impurities $i\neq j$, a $\rho^2\alpha^2$ mean field
contribution cancelled by
$\mean{\Omega }$, or a term containing two interactions with the
same impurity $i=j$, a $\rho\alpha^2$ term. It rewrites 
\bea\label{stop}
\mean{\delta\tilde{G}'^{(2)}_\beta(\vec r,\vec r\,')}&=& \begin{picture}(0,0)%
\epsfig{file=pertG2m.pstex}%
\end{picture}%
\setlength{\unitlength}{0.00087500in}%
\begingroup\makeatletter\ifx\SetFigFont\undefined
% extract first six characters in \fmtname
\def\x#1#2#3#4#5#6#7\relax{\def\x{#1#2#3#4#5#6}}%
\expandafter\x\fmtname xxxxxx\relax \def\y{splain}%
\ifx\x\y   % LaTeX or SliTeX?
\gdef\SetFigFont#1#2#3{%
  \ifnum #1<17\tiny\else \ifnum #1<20\small\else
  \ifnum #1<24\normalsize\else \ifnum #1<29\large\else
  \ifnum #1<34\Large\else \ifnum #1<41\LARGE\else
     \huge\fi\fi\fi\fi\fi\fi
  \csname #3\endcsname}%
\else
\gdef\SetFigFont#1#2#3{\begingroup
  \count@#1\relax \ifnum 25<\count@\count@25\fi
  \def\x{\endgroup\@setsize\SetFigFont{#2pt}}%
  \expandafter\x
    \csname \romannumeral\the\count@ pt\expandafter\endcsname
    \csname @\romannumeral\the\count@ pt\endcsname
  \csname #3\endcsname}%
\fi
\fi\endgroup
\begin{picture}(2115,1057)(226,-901)
\put(2341,-736){\makebox(0,0)[lb]{\smash{\SetFigFont{12}{14.4}{rm}$\vec r$}}}
\put(1756,-871){\makebox(0,0)[lb]{\smash{\SetFigFont{10}{12.0}{rm}$\alpha$}}}
\put(856,-871){\makebox(0,0)[lb]{\smash{\SetFigFont{10}{12.0}{rm}$\alpha$}}}
\put(1327, 48){\makebox(0,0)[lb]{\smash{\SetFigFont{10}{12.0}{rm}$\rho$}}}
\put(226,-736){\makebox(0,0)[lb]{\smash{\SetFigFont{12}{14.4}{rm}$\vec r\,'$}}}
\end{picture}

\\
&=& \pi\rho(i\alpha)^2
\int_0^\beta d\beta_1 \int_0^{\beta_1}d \beta_2 \int d\vec r_1 d\vec r_2\:
G^L_{\beta-\beta_1}(\vec r,\vec r_1)\frac{z_1-z_2}{\bar z_2-\bar z_1}
\nonumber\\
& &\hspace{3cm}
\Pi^{<L>}_{-} G^{<L>}_{\beta_1-\beta_2}(\vec r_1,\vec r_2)
\Pi^{<L>}_{ -} G^{<L>}_{\beta_2}(\vec r_2,\vec r\,')\nonumber
\eea
Using the algebra of the average Landau operators $\Pi^{<L>}_-$,
$\Pi^{<L>}_+$, $B^{<L>}_-$ and $B^{<L>}_+$, with 
\bea
B^{<L>}_-&=&-2i( \partial_{\bar z} - \frac{1}{2}\mean{\omega_c} z)\\
B^{<L>}_+&=&-2i( \partial_z        + \frac{1}{2}\mean{\omega_c} \bar z)
\eea
and the completness relations for the propagators, 
the  space integrations in (\ref{stop}) can be done
to obtain
\bea\mean
{\delta\tilde{ G}'^{(2)}_\beta (\vec r,\vec r\,')}
&=& -\pi\rho(i\alpha)^2\frac{\beta^2}{2}\Pi^{<L>}_+\Pi^{<L>}_- G^{<L>}_\beta(\vec r,\vec
r\,')\\
&=&-\alpha\mean{\omega_c}\beta^2
\frac{\partial}{\partial\beta}G^{<L>}_\beta (\vec r,\vec r\,')
\eea
Therefore, the correction to the average  partition function at order
$\alpha$ is
\begin{equation}\label{respert}
\delta Z_\beta=\alpha\frac{V\mean{\omega_c}}{\pi}
\frac{(\beta\mean{\omega_c})^2}{2\sinh^2(\beta\mean{\omega_c})}
\end{equation}
Accordingly, the correction to the average Landau density of states is
\be 
\delta\rho(E)= \alpha\mean{\omega_c}{\mean{\omega_c}\over \pi}{d^2\over dE^2}\sum_{n=0}^{\infty}
2(n+1)\mean{\omega_c}\delta(E-2(n+1)\mean{\omega_c})\ee

One can check that (\ref{respert}) reproduces the correction to the
mean field term in the one vortex problem, by simply taking the small
$\mean{\omega_c}$ limit in (\ref{respert})
\be \alpha\frac{V\mean{\omega_c}}{\pi}{1\over 2}={\alpha^2\over 2} V\rho\ee
which is nothing but the $\alpha^2$ term in (\ref{depletion'}) if one drops
the $\rho V=N$ factor.

Considering the conductivity,  we have in general to evaluate
\begin{eqnarray} \label{theend}
\sigma^-_{\beta}(t)&=&{\bigg\langle} i\theta(t)\frac{e^2}{V}
{1\over Z_{\beta}}
\int d\vec{r}\,d\vec{r}\,'
\Big( 
\Pi^{<L>}_-\tilde {G'}_{i t}(\vec{r},\vec{r}\,')
x' \tilde {G'}_{\beta-i t}(\vec{r}\,',\vec{r})
\nonumber\\
&& \hspace{5cm}-\ (i t\ \rightarrow\ i t+\beta)\Big){\bigg\rangle}
\end{eqnarray}
or its Fermi transform 
\begin{eqnarray}\label{final'}\sigma^-_{E_F}(t)&=&{\bigg\langle}i\theta(t)\frac{e^2}{V}
\int_{-\infty}^{\infty}\frac{dt'}{2i\pi}{e^{iE_Ft'}\over
t'-i\eta'}\int d\vec{r}\,d\vec{r}\,'
\Big( 
\Pi^{<L>}_-\tilde {G'}_{i t}(\vec{r},\vec{r}\,')
x' \tilde {G'}_{it'-i t}(\vec{r}\,',\vec{r})
\nonumber\\
&& \hspace{5cm}-\ ( t\ \rightarrow\ t+t')\Big){\bigg\rangle}
\end{eqnarray}
At first order in perturbation theory, one has to compute 

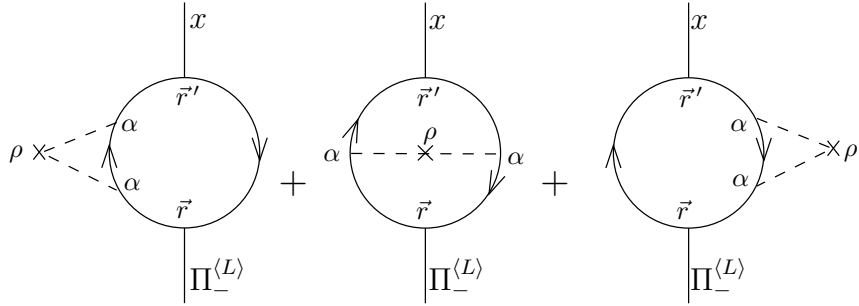
\vspace{1cm}
\begin{picture}(0,0)%
\epsfig{file=pertS2.pstex}%
\end{picture}%
\setlength{\unitlength}{0.00087500in}%
\begingroup\makeatletter\ifx\SetFigFont\undefined
% extract first six characters in \fmtname
\def\x#1#2#3#4#5#6#7\relax{\def\x{#1#2#3#4#5#6}}%
\expandafter\x\fmtname xxxxxx\relax \def\y{splain}%
\ifx\x\y   % LaTeX or SliTeX?
\gdef\SetFigFont#1#2#3{%
  \ifnum #1<17\tiny\else \ifnum #1<20\small\else
  \ifnum #1<24\normalsize\else \ifnum #1<29\large\else
  \ifnum #1<34\Large\else \ifnum #1<41\LARGE\else
     \huge\fi\fi\fi\fi\fi\fi
  \csname #3\endcsname}%
\else
\gdef\SetFigFont#1#2#3{\begingroup
  \count@#1\relax \ifnum 25<\count@\count@25\fi
  \def\x{\endgroup\@setsize\SetFigFont{#2pt}}%
  \expandafter\x
    \csname \romannumeral\the\count@ pt\expandafter\endcsname
    \csname @\romannumeral\the\count@ pt\endcsname
  \csname #3\endcsname}%
\fi
\fi\endgroup
\begin{picture}(5023,1825)(42,-1515)
\put(1048,-976){\makebox(0,0)[lb]{\smash{\SetFigFont{10}{12.0}{rm}$\vec r$}}}
\put(1127,163){\makebox(0,0)[lb]{\smash{\SetFigFont{12}{14.4}{rm}$x$}}}
\put(1666,-826){\makebox(0,0)[lb]{\smash{\SetFigFont{14}{16.8}{rm}$+$}}}
\put(2568,162){\makebox(0,0)[lb]{\smash{\SetFigFont{12}{14.4}{rm}$x$}}}
\put(3241,-826){\makebox(0,0)[lb]{\smash{\SetFigFont{14}{16.8}{rm}$+$}}}
\put(4143,162){\makebox(0,0)[lb]{\smash{\SetFigFont{12}{14.4}{rm}$x$}}}
\put(3042,-650){\makebox(0,0)[lb]{\smash{\SetFigFont{10}{12.0}{rm}$\alpha$}}}
\put(1934,-635){\makebox(0,0)[lb]{\smash{\SetFigFont{10}{12.0}{rm}$\alpha$}}}
\put(732,-811){\makebox(0,0)[lb]{\smash{\SetFigFont{10}{12.0}{rm}$\alpha$}}}
\put(717,-451){\makebox(0,0)[lb]{\smash{\SetFigFont{10}{12.0}{rm}$\alpha$}}}
\put( 42,-608){\makebox(0,0)[lb]{\smash{\SetFigFont{10}{12.0}{rm}$\rho$}}}
\put(5065,-602){\makebox(0,0)[lb]{\smash{\SetFigFont{10}{12.0}{rm}$\rho$}}}
\put(4394,-783){\makebox(0,0)[lb]{\smash{\SetFigFont{10}{12.0}{rm}$\alpha$}}}
\put(4383,-456){\makebox(0,0)[lb]{\smash{\SetFigFont{10}{12.0}{rm}$\alpha$}}}
\put(2533,-501){\makebox(0,0)[lb]{\smash{\SetFigFont{10}{12.0}{rm}$\rho$}}}
\put(1126,-1411){\makebox(0,0)[lb]{\smash{\SetFigFont{12}{14.4}{rm}$\Pi_-^{\mean{L}}$}}}
\put(2566,-1411){\makebox(0,0)[lb]{\smash{\SetFigFont{12}{14.4}{rm}$\Pi_-^{\mean{L}}$}}}
\put(4141,-1411){\makebox(0,0)[lb]{\smash{\SetFigFont{12}{14.4}{rm}$\Pi_-^{\mean{L}}$}}}
\put(4058,-984){\makebox(0,0)[lb]{\smash{\SetFigFont{10}{12.0}{rm}$\vec r$}}}
\put(4074,-309){\makebox(0,0)[lb]{\smash{\SetFigFont{10}{12.0}{rm}$\vec r\,'$}}}
\put(2495,-305){\makebox(0,0)[lb]{\smash{\SetFigFont{10}{12.0}{rm}$\vec r\,'$}}}
\put(2491,-984){\makebox(0,0)[lb]{\smash{\SetFigFont{10}{12.0}{rm}$\vec r$}}}
\put(1047,-298){\makebox(0,0)[lb]{\smash{\SetFigFont{10}{12.0}{rm}$\vec r\,'$}}}
\end{picture}

\vspace{1cm}

It is clearly a tedious task to evaluate  perturbatively (\ref{theend}),
already at first order in $\alpha$.
However, we will see below how to circumvent this computation.

\subsection{$\rho\alpha^n$ term: the one impurity case}

Before doing so, let us check that,
at leading order in $\alpha$, the  conductivity for the
one vortex case coincides with the low $\mean{\omega_c}$ expansion of the
Landau conductivity for the mean magnetic field. This should be so, since
we know that at  leading order $(\rho\alpha)^n$, the $n$ random magnetic
impurity system yields the Landau sytem (\ref{landaures}) 
at order  $\mean{\omega_c}^n$
\be \label{titi}
\re\sigma^{<L>}_{E_F}(\omega)|_{yx}=N(E_F) 
\frac{e^2}{V}\frac{2\mean{\omega_c}}{\omega^2}
\left( 1+\left(\frac{2\mean{\omega_c}}{\omega}\right)^2+
   \left(\frac{2\mean{\omega_c}}{\omega}\right)^4 +\cdots 
\right)
\ee
One has just to multiply the leading
term in $\alpha$ of the one vortex conductivity (\ref{oui}) by
$N$ to get the leading $\rho\alpha$ term 
\begin{equation}
\re\sigma_{E_F}(\omega)|_{yx}\APPROX{\alpha\to 0}
N(E_F)\frac{e^2}{V}2\pi\rho\alpha\frac{1}{\omega^2}=N(E_F)\frac{e^2}{V}
\,e\langle B\rangle\,\frac{1}{\omega^2}
\end{equation}
where the magnetic impurity density $\rho=N/V$, has factorized in $\mean{B}$.
One recovers the leading term in the expansion (\ref{titi}) of the conductivity of the Landau
problem in the low mean magnetic field limit $\mean{\omega_c}/\omega\to 0$.
If, in the single impurity case, the conductivity is
divergent at small $\beta\omega$, in the random magnetic impurity case, this
divergence is nothing but a manifestation of the small $\mean{\omega_c}/\omega$ expansion
of the mean Landau conductivity.
On the other hand, considering the longitudinal conductivity for the Landau
problem (\ref{sans}), its small $\mean{\omega_c}$ expansion starts at order
$\mean{\omega_c}^2$, a $\rho^2$ behavior which
cannot be obtained from the single vortex problem (\ref{non}).

\subsection{ $\rho^n\alpha^{n+1}$ terms: the one vortex +
a homogeneous magnetic field }
As advocated above, perturbation theory with (\ref{extract}) at first order
in $\alpha$
amounts to consider only the one 
impurity diagrams. It follows that, at this order, one can map the
magnetic impurity  perturbative problem
on the exact one vortex + a homogeneous mean magnetic field problem. 
Consider again the Hamiltonian (\ref{vbd}) for the  one vortex + $\mean{B}$
problem
\begin{equation}\label{vbdbis}
H_{d}=\frac{1}{2}\left(\vec{p}-\mean{\omega_c} \vec k\times\vec r-
\alpha \frac{\vec{k}\times\vec{r}}{r^2}\right)^2
+(\pi\alpha\delta(\vec r)+\mean{\omega_c})\ee
and redefine the wavefunctions in order to extract only the short distance
behaviour of
the wavefunctions at the origin due to the vortex 
\bea
U_d(\vec r)=r^{\alpha}
\eea
It follows that
\bea
U_d^{-1}\Pi_+ U_d&=&\Pi^{<L>}_+ - 2i\alpha{1\over {\bar z}} \\
U_d^{-1}\Pi_- U_d&=&\Pi^{<L>}_-
\eea

The $\tilde{H}_d$ Hamiltonian reads
\be \label{HBVtd}
\tilde{H}_d={1\over 2}\Pi^{<L>}_+\Pi^{<L>}_- - i
\alpha{1\over {\bar z}}\Pi^{<L>}_-
\ee
Comparing the Hamiltonian (\ref{extract}) with (\ref{HBVtd}), one sees
that they
can
be identified  if, leaving asides the contribution of the average potential
$\mean{\Omega}$,  one restricts (\ref{extract}) to one impurity located
at the origin. This
shows in particular that, apart for a factor $N$, the second order
$\alpha^2$ computation for the one vortex +$\mean{B}$  problem is
identical to the first order
$\rho\alpha^2\simeq \mean{\omega_c}\alpha$ correction to the mean field in the magnetic impurity
problem.  
One can check explicitly these considerations on the partition function.
From (\ref{ZBVd}), the $\alpha^2$ term reads 
\be
\delta Z_\beta^{(2)}(\mean{B},\alpha)
=\alpha^2\frac{(\beta\mean{\omega_c})^2}{2\sinh^2(\beta\mean{\omega_c})}
\ee
Multiplying it by $N$ reproduces the perturbative result
(\ref{respert}) of the magnetic impurity problem at order $\alpha$.

\subsection{Perturbative Hall conductivity}

We now  go back to the magnetic impurity problem, and instead of computing
perturbatively (\ref{theend}), we rather consider
in (\ref{BVcond})  the $\alpha^2$ correction multiplied
by $N$. One gets the conductivity for the magnetic impurity problem at first
order in $\alpha$
\bea
\sigma^-_{E_F}(\omega=0)&=&
\int_{-\infty}^{+\infty}\frac{dt'}{2i\pi}\frac{e^{iE_Ft'}}{t'-i\eta'}
\frac{i e^2}{2V\mean{\omega_c}}\Big\{
  Z_{it'} + \alpha\mean{\omega_c} it'\: Z^{<L>}_{it'}+\cdots 
\Big\} \\
&=& \frac{i e^2}{2V\mean{\omega_c}}\left\{
   N(E_F)+\alpha\mean{\omega_c}\frac{d N^{<L>}(E_F)}{d E_F}+\cdots 
\right\}
\eea
where $Z^{<L>}_\beta$ and $N^{<L>}(E_{F})$ are respectively the
partition function and the number of electrons for the average Landau problem.
$N^{<L>}(E_{F})$ is nothing but the mean field contribution to the actual
number of
electrons in the system $N(E_F)=N^{<L>}(E_F)+\alpha N^{(1)}({E_F})+
\alpha^2 N^{(2)}({E_F})+\cdots$ where the $ N^{(n)}$'s are
functions of $\mean{\omega_c}$ and $E_F$ only.
Then, one can rewrite
\be\label{ouf}
\sigma^-_{E_F}(\omega=0)=
\frac{i e^2}{2V\mean{\omega_c}}N(E_F+\alpha\mean{\omega_c})
\ee
which is still valid at  first order in $\alpha$.
The Hall conductivity is thus, at this order,
\be\label{smooth}
\re\sigma_{E_F}(\omega=0)|_{yx} =
-N(E_F+\alpha\mean{\omega_c}){1\over V}\frac{e}{\mean{B}}\ee
a quite simple result (eventhough higher order perturbation theory might
change this situation).

Also, from (\ref{soBV}), one  finds no correction for the longitudinal
conductivity, which means that at this order, (\ref{sans}) is not affected by
disorder.
\section{Heuristic considerations and open questions}

Let us consider the experimental  situation where the magnetic field
$\mean{B}$ is kept fixed, and where the electron density varies. The Hall
conductivity (\ref{smooth}) is, basically, $N(E_F+\alpha\mean{\omega_c})$ as a function of
$N(E_F)$.
 In~\cite{nous},
the average density of states ${\rho(E)}$  was
estimated on analytical (approximation) and
numerical grounds. In Figure \ref{r01}, ${\rho(E)}$ is given for $\alpha=0.01$. This curve
has essentially an $\alpha$ behavior, with broadened Landau levels having a
width
of order  $\alpha\mean{\omega_c}$, a height of order  $1/\alpha$, such
that the total number
of quantum states per unit surface in a broadened Landau level,
$\mean{\omega_c}/\pi$, is conserved
whatever the disorder is.

In Figure \ref{syx}, the Hall conductivity (\ref{smooth})
is displayed (full line) as a function of the number of electrons
$N(E_F)$, while keeping the mean magnetic field fixed. The Hall conductivity 
has oscillations close to
the classical
straight line of the average Landau conductivity. We believe that
the oscillations are smoothed
because
of the approximation made in (\ref{ouf}) and in
${\rho(E)}$, and consequently in $N(E_F)$. In fact ${\rho(E)}$ is
too broad to get a stepwise Hall conductivity. On the other hand, we also
considered that ${\rho(E)}$ at small $\alpha$ is the same object as
${\rho(E)}$ at first order in $\alpha$. But the expansion of the
density of states around the mean field density of states is surely not
analytical. To conclude, the curve given in Figure~\ref{syx} should only be viewed
as a crude first step
towards a more complete computation. 
It is not
impossible that higher order corrections, or a more general argument,
will yield a behavior closer to the
experimental curve (dashed lines in Figure \ref{syx}, from~\cite{jansen}).

What kind of physical picture could support the model presented above?
When the magnetic field is
very strong,  $\alpha\simeq 1/2$, the system is totally
disordered, and cannot possibly transport any current.  This is
consistent with the symmetry argument which indicates that at $
\alpha=1/2$ the transverse conductivity has to vanish.
On the other hand, when $\mean{B}$ decreases, Landau level
oscillations appear, implying that some states conduct, thus the 
appearance of plateaus. When
$\mean{B}$ becomes smaller and smaller, i.e. $\alpha\to 0$, 
a pure
Landau system is reached with a classical Hall conductivity and no plateau at all.

Clearly, no edge or finite size effects  have been considered  whatsoever,
and no interaction between electrons either. Also, the thermal average has 
forbidden
any precise information on the possible localisation of  quantum states
in the random magnetic impurity distribution.
 
It would be certainly rewarding to know  how to extract from (\ref{theend})
a more global (non perturbative) information,
which would, like the Thouless topological argument~\cite{thouless}, 
tell us for sure that, when considering 
the Fermi gas at zero temperature and Fermi
energy $E_F$, the correction terms always oscillate around the classical
straight line.

\section*{Acknowledgments} 

We are  pleased to acknowledge useful conversations with Eric Akkermans,
Alain Comtet,
Cyril Furtlehner, Patricio Leboeuf and
Amaury Mouchet. One of us (S.O.) acknowledges useful conversations with Jan
Myrheim.

\begin{figure}[!b]
\begin{center}
\input{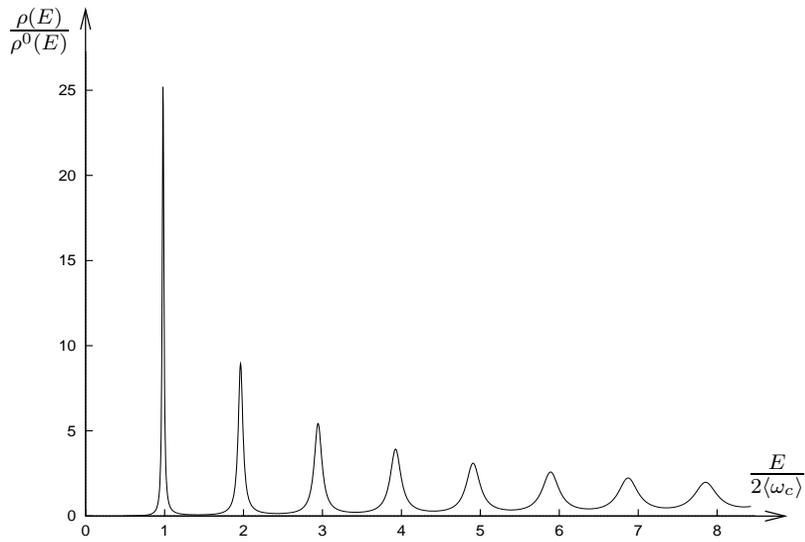}
\caption{\label{r01} Average density of states of the random magnetic
impurity model for $\alpha=0.01$}
\end{center}
\end{figure}

\begin{figure}
\begin{center}
\input{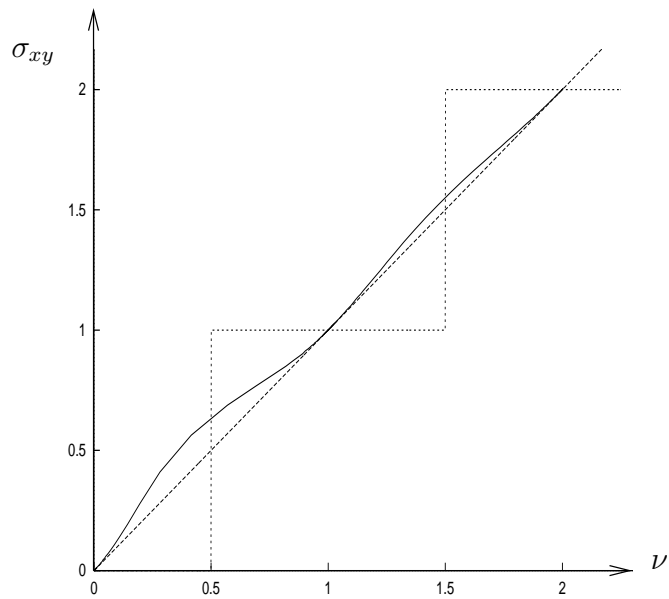}
\caption{\label{syx} Hall conductivity in unit of $e^2/h$
of the random magnetic impurity model
at first order in $\alpha$
for $\alpha=0.01$ as a function of the filling factor $\nu={N(E_F)\over
V}{h\over e\mean{B}}$: straight line = classical result, steps = experimental
Integer Quantum Hall Effect (schematic), full  line = perturbative result}
\end{center}
\end{figure}

\end{document}